\documentclass[11pt,a4paper]{article}

\usepackage{latexsym,amsfonts,amssymb,amsmath}

\newcommand{\mR}{\mathbb{R}}
\newcommand{\PP}{\mathbb P}
\newcommand{\QQ}{\mathbb Q}
\newcommand{\MM}{\mathbb M}
\newcommand{\TT}{\mathbb T}
\newcommand{\e}{\mathrm{e}}

\newcommand{\kp}{$ k \cdot p $ }
\newcommand{\Bf}{\mathit{bf}}
\newcommand{\Br}{\mathit{br}}
\newcommand{\Is}{\mathit{is}}
\newcommand{\vctr}[1]{\mathaccent"017E {#1} }
\newcommand{\abs}[1]{{\vert {#1} \vert}}

\newcommand{\pt}{\partial}
\newcommand{\momentou}{\mathbf u}
\newcommand{\vF}{\mathbf F}
\newcommand{\vU}{\mathbf U}
\newcommand{\vp}{\mathbf p}
\newcommand{\vx}{\mathbf x}
\newcommand{\vy}{\mathbf y}
\newcommand{\vk}{\mathbf k}
\newcommand{\vu}{\mathbf u}
\newcommand{\vv}{\mathbf v}

\newcommand{\valpha}{\text{\boldmath$\alpha$}}

\newcommand{\vxi}{\text{\boldmath$\xi$}}
\newcommand{\vK}{\mathbf K}
\newcommand{\vB}{\mathbf B}

\newcommand{\vq}{\mathbf q}
\DeclareMathOperator{\Tr}{Tr}

\newtheorem{theorem}{Theorem}
\newtheorem{remark}{Remark}

\begin{document}

\begin{center}

{\bf \huge Semiclassical hydrodynamics \\  of a quantum  Kane model 
\\[8pt] for semiconductors}

\par 
\vspace{20pt}
{\sc \large L.~Barletti, G.~Borgioli, G.~Frosali}
\par 
\vspace{10pt}
Dipartimento di Matematica e Informatica ``U.Dini'', Firenze, Italy
\par 
\vspace{5pt}
{\tt \small luigi.barletti@unifi.it, giovanni.borgioli@unifi.it, giovanni.frosali@unifi.it}

\end{center}

\vspace{8pt}

\begin{abstract}
In this paper we derive a semiclassical hydrodynamic system
for electron densities and currents in the two energy bands of a
semiconductor. We use the semiclassical Wigner equation with a
$k\cdot p$ Hamiltonian and a BGK dissipative term to construct the
first two moment equations. The closure of the moment system is
obtained using the Maximum Entropy Principle, by minimizing a
Gibbs free-energy functional under suitable constraints. We prove that the constraint equations can be uniquely solved,
i.e. that the local equilibrium state can be parametrized by the density and velocity field. Some BGK-like models are
proposed to mimic the quantum interband migration.
\end{abstract}



\section{Introduction}
Description of the charge carriers dynamics in semiconductor devices
is certainly a severe task, especially if one wishes to keep
together a rigorous (and complete, whenever possible) physical
picture with a final result (set of equations) simple enough for the
numerical implementation. Hydrodynamic approach is an excellent
compromise between the two requirements. Our aim is the construction
of hydrodynamic equations for the electron dynamics, by means of moment method, starting from the pseudo-kinetic formulation
of quantum mechanics in terms of Wigner functions.
The physical framework adopted in this paper is based on the so
called $k\cdot p$ method, \cite{Kane66,Wenckebach99}, a simple model
for the description of
charge transport in a semiconductor with two available energy bands.

The $k\cdot p$ Hamiltonian has been widely studied and employed in literature (see
for instance the review \cite{BarFroMor2014}). In particular, it has been exploited in \cite{BarlettiFrosali10,BarlettiFrosali12}
to derive a semi-classical two-band diffusive model, with weak or strong external fields.

The rigorous derivation of the $k\cdot p$ Hamiltonian
from the complete Hamiltonian of an electron in a periodic potential, under a suitable homogenization
scaling, is based on the concept of envelope functions and can be found in \cite{BarlettiBenAbdallah11}.
The result is a  $2 \times 2$ matrix Hamiltonian, which means that electrons in the \kp description are pseudo-spinors (the pseudo-spin being
related to the two energy bands).
A fully-quantum treatment based on the \kp method leads to non-parabolic intraband dynamics as well as to interband quantum transitions.
However, in the present semiclassical treatment, the latter aspect is lost. Nevertheless, the non-parabolic dynamics is still present
and leads to non-trivial fluid models.
\par
The semiclassical kinetic equations, that we need to get the hydrodynamic model, can be naturally expressed in terms of Wigner functions,
describing statistical states of electrons in terms of quasi-distributions in phase-space.
Due to pseudo-spin, the standard scalar Wigner function has to be substituted by a matrix-valued Wigner function.
Such a matrix can be projected on the two energy subspaces, thus obtaining two distributions of electrons, corresponding to the two energy bands.
Then, the macroscopic fluid quantities can be obtained by taking moments of the band-projected Wigner function, which have the physical
meaning of densities $n_\pm$ and  velocity field $\vu_\pm$, where the subscript $\pm$ means $+$, the upper band, and $-$, the lower band (see Eqs. \eqref{momento_n}
and \eqref{momento_u}.
The Wigner formalism, moreover, permits the introduction of a well justified BGK term (see \cite{Arnold96,DegondRinghofer03}]) which takes in account the interaction
 phenomena leading to a local equilibrium relaxation.
 Thanks to this relaxation mechanism we can assume that, in a time-scale larger that the relaxation time, the system is in a local equilibrium
 state. The latter is chosen according to the Maximum Entropy Principle (MEP), i.e.\ as the most probable microscopic state, given the observed
 macroscopic moments  $n_\pm$ and  $\vu_\pm$. This strategy, as usual, provides a closure of the moment equations.
%

The paper is organized in the following way:
in section 2 we present the $k\cdot p$ Hamiltonian. The
presence of the two bands is treated introducing a pseudo-spinorial
formulation via a representation on the Pauli matrices basis.
%
In section 3 we deduce the Wigner-BGK equations for our
model. The Wigner matrix is decomposed in its scalar part $w_0$ and
its pseudo-spinorial part $\vctr w$. $\vctr w$ is further split in a
part parallel to the direction of the pseudo-spinorial part of the
Hamiltonian, $w_S$, and a part orthogonal to it, $\vctr{w}_\perp$.
This representation discovers itself useful in the evaluation of the moments for the Wigner
equation, since the contribution
of $\vctr{w}_\perp$  vanishes.
%
In Section 4 we deduce the moment equations of zeroth and first order, where appear the tensors $\mathbb{P}_\pm$ and $\mathbb{Q}_\pm$, which can
be interpreted as  the pressure and effective-mass tensors.
%
In Section 5 the application of the MEP implies
that these tensors depend on two Lagrange multipliers, a scalar one, $A_\pm$, and a vector one, $\mathbf{B}_\pm$. The closure of the moment equations requests
the study of the dependence of the tensors on the macroscopic quantities, $n_\pm$, the numerical density and $\mathbf{u}_\pm$, the velocity field.
In Theorem 1 we prove that $\mathbf{B}_\pm$ (and $A_\pm$, as a consequence) is a smooth globally invertible function of the macroscopic quantities.

Since in semiclassical limit the quantum interference terms between the two bands disappear, in Section 6 we examine some models that enable the reintroduction of
this aspect. We propose there three different BGK-like terms which satisfy this condition.


\section{The \kp model}
\label{S2}
The simplest possible description of an electron in a semiconductor crystal
with two energy bands (e.\ g.\ ``valence'' and  ``conduction'') is obtained from a periodic Hamiltonian by
means of the \kp method \cite{Kane66, Wenckebach99} and consists of a $2\times2$ Hamiltonian
of the following form:
\begin{equation}
\label{KpHam}
 H =
  \begin{pmatrix}
   -\frac{\hbar^2}{2m}\Delta + E_g/2     &  - \frac{\hbar^2}{m}\,\vK\cdot\nabla
  \\[12pt]
   \frac{\hbar^2}{m}\,\vK\cdot\nabla   &  -\frac{\hbar^2}{2m}\Delta - E_g/2
  \end{pmatrix}.
\end{equation}
Here, $E_g$ is the band-gap and $\vK = (K_1,K_2,K_3)$ is the matrix element of the gradient
operator between the Bloch functions $b_\pm$ of the upper ($+$) and lower ($-$) bands,
evaluated at zero pseudo-momentum:
$$
  \vK = \int\limits_\text{\it lattice cell} \overline b_+(\vx)\, \nabla b_-(\vx)\,d\vx\,,
$$
$\hbar$ is Planck's constant over $2\pi$ and $m$ is the electron mass.
The \kp model has to be completed by adding an ``external'' potential term $qV$ (where $q>0$ denotes
the elementary charge), accounting for all electric fields except the crystal one.
The electric potential $V(\vx)$ can be either fixed or self-consistently given by
a Poisson equation.
\par
The \kp Hamiltonian $H$ is the quantization of the classical matrix-valued symbol
\begin{equation}
\label{KpHamSymb}
  h(\vp) =
  \begin{pmatrix}
   \frac{p^2}{2m} + E_g/2     &  - i\frac{\hbar}{m} \vK \cdot \vp
  \\[12pt]
   i\frac{\hbar}{m} \vK \cdot \vp   &  \frac{p^2}{2m} -E_g/2
  \end{pmatrix},
\end{equation}
where $p = \abs{\vp}$.
\par
In this paper we make the choice to decompose any $2\times 2$ complex matrix in the basis of Pauli
matrices
$$
 \sigma_0 = \begin{pmatrix} 1 & 0 \\ 0 & 1   \end{pmatrix},
\quad
 \sigma_1 = \begin{pmatrix} 0 & 1 \\ 1 & 0   \end{pmatrix},
\quad
 \sigma_2 = \begin{pmatrix} 0 &-i \\ i & 0   \end{pmatrix},
\quad
  \sigma_3 =  \begin{pmatrix} 1 & 0 \\ 0 &-1   \end{pmatrix},
$$
(the coefficients of the decomposition will be real if the matrix is hermitian). The operators
$\sigma_1$, $\sigma_2$, $\sigma_3$ are called ``pseudo-spin components'' in this context.
Putting
\begin{equation}
\label{AlphaGammaDef}
\valpha =  (\alpha_1,\alpha_2,\alpha_3) := \frac{\hbar}{m}\, \vK
\quad \text{and} \quad
\gamma := E_g/2,
\end{equation}
we can write
\begin{equation}
\label{KpSymb}
  h(\vp) =
  \frac{p^2}{2m}\,\sigma_0 + {\valpha \cdot \vp} \,\sigma_2 + \gamma \,\sigma_3
  = h_0(\vp)\sigma_0 + \vctr{h}(\vp) \cdot \vctr{\sigma},
\end{equation}
where
\begin{equation*}
 h_0(\vp) = \frac{p^2}{2m}, \qquad \vctr{h}(\vp) = (0,\valpha\cdot \vp, \gamma),
\end{equation*}
and, as usual, $\vctr{\sigma} = (\sigma_1,\sigma_2,\sigma_3)$ is the formal vector of Pauli matrices.
%
Here and in the following we adopt the arrow notation for three-vectors, such as $\vctr{h}(\vp)$, that are the
pseudo-spinorial part of the Pauli coefficients.
Instead, we do not use the arrow notation for ``cartesian'' three-vectors such as $\vx$, $\vp$, $\vK$, $\valpha$,  etc.
%
The dispersion relation for the free Hamiltonian $H$ is easily obtained  by
computing the ($\vp$-dependent) eigencouples of the symbol $h(\vp)$.
This yields to the energy bands
\begin{equation}
\label{Bands}
  E_\pm(\vp) = \frac{p^2}{2m} \pm \sqrt{(\valpha\cdot \vp)^2 + \gamma^2}
 = \frac{p^2}{2m} \pm \abs{\vctr{h}(\vp)}
\end{equation}
and to the corresponding normalized energy eigenvectors
\begin{equation}
\label{psipDef}
 \psi^p_\pm = \frac{1}{\sqrt{2(1 \pm \nu_3(\vp))}}
 \begin{pmatrix} \nu_3(\vp) \pm 1 \\ \nu_1(\vp) + i\nu_2(\vp)\end{pmatrix},
\end{equation}
where we have introduced
\begin{equation}
\label{nuDef}
  \vctr{\nu}(\vp) = (\nu_1(\vp), \nu_2(\vp), \nu_3(\vp)) = \frac{\vctr{h} (\vp)}{\abs{\vctr{h}(\vp)}}
   = \frac{(0,\valpha\cdot \vp, \gamma)}{\sqrt{(\valpha\cdot \vp)^2 + \gamma^2}} .
\end{equation}
The two eigenprojections $P_\pm(\vp)$, that we call band-projections, are therefore given by
\begin{equation}
\label{Pdef}
  P_\pm(\vp) = \psi_\pm^p \otimes \psi_\pm^p = \frac{1}{2} (\sigma_0 \pm \vctr{\nu}(\vp) \cdot \vctr{\sigma})
\end{equation}
and we can clearly write
\begin{equation}
\label{SpecDec}
  h(\vp) =  E_+(\vp) P_+(\vp) + E_-(\vp) P_-(\vp).
\end{equation}
Important quantities associated to the energy bands are the semiclassical velocities $ \vv_\pm$
\begin{equation}
\label{velocity}
 \vv_\pm = \nabla_\vp  E_\pm(\vp)
 = \frac{\vp}{m} \pm  \frac{\valpha \cdot \vp}{ \sqrt{(\valpha\cdot \vp)^2 + \gamma^2}} \valpha
 = \frac{\vp}{m} \pm
\nu_2 \valpha
\end{equation}
and the  effective-mass tensor $\MM_\pm(\vp)$ defined by \cite{BarlettiBenAbdallah11}
\begin{equation}
\label{effmass}
\MM_\pm^{-1}(\vp) = \nabla_\vp \otimes\nabla_\vp E_\pm(\vp)
= \frac{1}{m}{\mathbb I} \pm \frac{\gamma^2 \valpha\otimes\valpha}{\left((\valpha\cdot \vp)^2 + \gamma^2\right)^{3/2}}.
\end{equation}
where $\mathbb I$ is the identity matrix.
\section{Wigner-BGK equations for the \kp model}
\label{S3}
Let $\rho_{ij}(\vx,\vy,t)$, $1 \leq i,j \leq 3$, be the density matrix describing the quantum statistical state of electrons
with Hamiltonian \eqref{KpHam}.
The corresponding kinetic-like description is provided by the Wigner matrix $w_{ij}(\vx,\vp,t)$ defined by
\cite{Wigner32,ZachosEtAl05,BarlettiFrosali10}
\begin{equation}
\label{WigDef}
  w_{ij}(\vx,\vp,t) = \frac{1}{(2\pi\hbar)^{3/2}}\int_{\mR^3}
  \rho_{ij}\left( \vx + \frac{\vxi}{2},  \vx - \frac{\vxi}{2},t \right) \e^{-i\vp \cdot \vxi/\hbar}  d\vxi.
\end{equation}
The Wigner matrix $w = (w_{ij})$ is hermitian,
\begin{equation*}
 w(\vx,\vp,t) = w^*(\vx,\vp,t),
 \end{equation*}
 and, consequently, its Pauli representation
\begin{equation}
\label{WigPauli}
  w = w_0 \sigma_0 + \vctr{w} \cdot \vctr{\sigma}, \qquad \vctr{w} = (w_1,w_2,w_3)
\end{equation}
has real components $w_k(\vx,\vp,t)$, $0\leq k\leq 3$.
\par
Considering $P_\pm$ and $\vctr{\nu}$, as defined in \eqref{Pdef}  and \eqref{nuDef}, the two scalar functions
\begin{equation}
\label{wpm}
  w_\pm = \Tr(P_\pm w) =  w_0 \pm \vctr{\nu} \cdot \vctr{w}
\end{equation}
can be semi-classically interpreted as the phase-space distributions of electrons in the two energy bands $E_\pm$
\cite{BarlettiFrosali10} and will play a central role in the following.
Moreover, if $w_s = \vctr{\nu}\cdot\vctr{w}$, we have the obvious relations
\begin{equation}
\label{relations}
  w_\pm = w_0 \pm w_s,
  \qquad
  w_0 = \frac{w_++w_-}{2},
  \qquad
  w_s = \frac{w_+-w_-}{2},
\end{equation}
and $w_s$  has therefore the meaning of ``band polarization''.
It will be convenient, moreover, to introduce a notation for the perpendicular part of $\vctr{w}$ with respect to $\vctr{\nu}$
by putting
\begin{equation}
\label{vector1}
\vctr{w}=w_s\vctr{\nu}+\vctr{w}_\bot.
\end{equation}
Assume now that the dynamics of the density matrix $\rho(\vx,\vy,t)$ is given by the
von Neumann equation (Schr\"odinger equation for mixed states)
\begin{equation*}
  i\hbar\,\frac{\pt \rho}{\pt t} = (H_\vx-H_\vy)\rho +\left(V(\vx) - V(\vy)\right)\sigma_0\rho,
\end{equation*}
where $H_\vx$ and $H_\vy$ denote the \kp Hamiltonian \eqref{KpHam} acting, respectively, on the $\vx$ and
$\vy$ variables, and $V$ is an external and/or self-consistent electric field.
Then, using \eqref{WigDef} and \eqref{WigPauli}, it is not difficult to prove that, up to terms of order $\hbar^2$, the evolution equations for the time dependent Pauli-Wigner functions are the following
\begin{equation}
\label{WE}
\left\{
\begin{aligned}
 &\frac{\pt w_0}{\pt t}  + \frac{\vp}{m}\cdot\nabla_\vx w_0+ \vF \cdot \nabla_\vp w_0
 + \valpha\cdot\nabla_\vx w_2 = 0,
\\[4pt]
 &\frac{\pt \vctr{w}}{ \pt_t}+ \frac{\vp}{m}\cdot\nabla_\vx  \vctr{w} + \vF \cdot \nabla_\vp \vctr{w}
  + \valpha\cdot\nabla_\vx w_0\,\vctr{e}_2 -  \frac{2}{\hbar}\, \vctr{h}(\vp) \times \vctr{w} =  0.
\end{aligned}
\right.
\end{equation}
Here, $\vctr{h}(\vp) = (0,\valpha\cdot \vp,\gamma)$, $\vctr{e}_2 = (0,1,0)$
and $\vF = - \nabla V$ denotes the external force corresponding to the electric potential $V$.
\par
In order to supplement system  \eqref{WE}, which describes a conservative Hamiltonian dynamics,
with a collisional mechanism, we insert a BGK (Bhatnaghar-Gross-Krook)  collisional relaxation-time term.
This term mimics the collisions  that  force the system towards a local equilibrium and it is characterized
by the relaxation time $\tau_c$,  which is assumed to be the same constant for all components.
The system, which will be referred to as ``Wigner-BGK'' (WBGK) equations, takes the new form
\begin{equation}
\label{WBGK1}
\left\{
\begin{aligned}
 &\frac{\pt w_0}{\pt t}  + \frac{\vp}{m}\cdot\nabla_\vx w_0+ \vF \cdot \nabla_\vp w_0
 + \valpha\cdot\nabla_\vx w_2 =  \frac{g_0 - w_0}{\tau_c},
\\[4pt]
 &\frac{\pt \vctr{w}}{ \pt t}+ \frac{\vp}{m}\cdot\nabla_\vx  \vctr{w} + \vF \cdot \nabla_\vp \vctr{w}
  + \valpha\cdot\nabla_\vx w_0\,\vctr{e}_2 -  \frac{2}{\hbar}\, \vctr{h}(\vp) \times \vctr{w}
  =   \frac{\vctr{g} - \vctr{w}}{\tau_c},
\end{aligned}
\right.
\end{equation}
where $g = g_0\sigma_0 + \vctr{g} \cdot \vctr{\sigma}$ is
a local-equilibrium Wigner matrix that will be specified later on.
\par
We now extract from Eq.\ \eqref{WBGK1}, equations for the band distributions $w_+$ and $w_-$ (see definition
\eqref{wpm}).
For this purpose we introduce the orthonormal basis $(\vctr{n}_1,\vctr{n}_2,\vctr{\nu})$, where
$\vctr{n}_1\equiv \vctr{e}_1 = (1,0,0)$ and $\vctr{n}_2$ is chosen such that
$\vctr{n}_1\times\vctr{n}_2=\vctr{\nu}$.
Using the decomposition $\vctr{w}=w_s\vctr{\nu}+\vctr{w}_\bot$ (see (\eqref{vector1})
and taking account that
$w_2 = w_s\vctr{\nu} \cdot \vctr{e}_2 + \vctr{w}_\bot \cdot \vctr{e}_2$, with
\begin{equation}
\label{baseedue}
\vctr{e}_2 =  \frac{\valpha \cdot \vp}{ \sqrt{(\valpha\cdot \vp)^2 + \gamma^2}} \vctr{\nu}
+ \frac{\gamma}{ \sqrt{(\valpha\cdot \vp)^2 + \gamma^2}} \vctr{n}_2,
\end{equation}
we rewrite the first of equations \eqref{WBGK1} as
\begin{multline}
\label{w_zero}
\frac{\pt w_0}{\pt t }  + \frac{\vp}{m} \cdot\nabla_\vx w_0 + \vF \cdot \nabla_\vp  w_0 +
 \frac{\valpha\cdot \vp}{\sqrt{(\valpha\cdot \vp)^2 + \gamma^2}} \valpha \cdot \nabla_\vx w_s
 \\
+ \frac{\gamma}{\sqrt{(\valpha\cdot \vp)^2 + \gamma^2}} \vctr{n}_2 \cdot \left(  \valpha \cdot \nabla_\vx \vctr{w}_\perp \right)
 =  \frac{g_0 - w_0}{\tau_c}.
\end{multline}
Concerning the second of equations \eqref{WBGK1}, using  again \eqref{baseedue}, we have
\begin{multline*}
   \frac{\pt}{\pt t }\left( w_s \vctr{\nu} +   \vctr{w}_\perp\right)
    +  \frac{\vp}{m}  \cdot\nabla_\vx\left( w_s \vctr{\nu} +   \vctr{w}_\perp\right)
    +  \frac{\valpha\cdot \vp\, \vctr{\nu} + \gamma\, \vctr{n}_2}{\sqrt{(\valpha\cdot \vp)^2 + \gamma^2}}
    \valpha\cdot\nabla_\vx w_0
   \\
    +  \vctr{\nu}\,\vF \cdot \nabla_\vp  w_s
    +   \left(\vF \cdot \nabla_\vp  \vctr{\nu}\right) w_s
    +  \vF \cdot \nabla_\vp \vctr{w}_\perp
  \\
  =   \frac{2}{\hbar}\, \vctr{h}(\vp) \times {\vctr w}_\perp
    + \frac{{g_s} - {w_s}}{\tau_c} \vctr{\nu} +  \frac{\vctr{g}_\perp - \vctr{w}_\perp}{\tau_c}.
\end{multline*}
Decomposing this equation in the parallel and perpendicular parts with respect to $\vctr{\nu}$,
and using $\vctr{\nu}\cdot \left(\vF\cdot\nabla_\vp \vctr{\nu}\right) = 0$, we obtain an equation for $w_s$:
\begin{multline}
\label{w_s}
 \frac{\pt w_s}{\pt t }   +  \frac{\vp}{m}  \cdot\nabla_\vx w_s  +  \vF \cdot \nabla_\vp w_s
 +  \frac{\valpha\cdot \vp}{\sqrt{(\valpha\cdot \vp)^2 + \gamma^2}}  \valpha\cdot\nabla_\vx w_0
 \\
+ \vctr{\nu}\cdot \left( \vF \cdot \nabla_\vp\vctr{w}_\perp\right)
=  \frac{{g_s} - {w_s}}{\tau_c},
\end{multline}
and  an equation for
$\vctr{w}_\perp$:
\begin{multline}
\label{w_p}
\frac{\pt  \vctr{w}_\perp}{\pt t } +  \frac{\vp}{m}  \cdot\nabla_\vx {\vctr w}_\perp
+(\vF\cdot\nabla_\vp)w_s
+\left( \vF \cdot \nabla_\vp \vctr{w}_\perp \right)_\perp
\\
+ \frac{\gamma\, \vctr{n}_2}{ \sqrt{(\valpha\cdot \vp)^2 + \gamma^2}}  \valpha\cdot\nabla_\vx w_0
= \frac{2}{\hbar}\, \vctr{h}(\vp) \times {\vctr w}_\perp + \frac{\vctr{g}_\perp - \vctr{w}_\perp}{\tau_c},
\end{multline}
(which will not be used in the following).
Then, recalling \eqref{relations} and \eqref{velocity},  equations for $w_+$ and $w_-$ are now readily obtained from \eqref{w_zero} and \eqref{w_s}:
\begin{multline}
\label{sum3}
 \frac{\pt w_\pm}{\pt t }  + \vv_\pm
 \cdot\nabla_\vx w_\pm  +   \vF \cdot \nabla_\vp  w_\pm   +
\frac{\gamma}{\sqrt{(\valpha\cdot \vp)^2 + \gamma^2}} \vctr{n}_2 \cdot \left(  \valpha \cdot \nabla_\vx \vctr{w}_\perp \right)
\\
 \pm \vctr{\nu}\cdot \left( \vF \cdot \nabla_\vp \vctr{w}_\perp\right)
 =  \frac{g_\pm - w_\pm}{\tau_c}.
\end{multline}

\section{Moment equations and entropy closure}
\label{S4}
The local equilibrium Wigner matrix $g = g_0 \sigma_0 + {\vctr g} \cdot {\vctr \sigma}$ is given by the
MEP and is, therefore, the maximizer of a suitable entropy functional (which depends
on the particle statistics) under the constraint of given macroscopic moments \cite{LaRosa09, Wu97}.
We make the following assumptions:
\begin{enumerate}
\item
the system is in thermal equilibrium at constant temperature $T>0$ (e.g.\ with a phonon bath);
\item
the electron statistics is well approximated by Maxwell-Boltzmann distribution (in the semiclassical approach);
\item
the observed macroscopic moments are the densities
\begin{equation}
\label{momento_n}
  n_\pm(\vx,t) = \int_{\mR^3} w_\pm(\vx,\vp,t)\,d\vp
\end{equation}
 and the velocity field
\begin{equation}
\label{momento_u}
  \vu_\pm(\vx,t) = \frac{1}{n_\pm(\vx,t)} \int_{\mR^3} \vv_\pm(\vp)\,w_\pm(\vx,\vp,t)\,d\vp
\end{equation}
of the electrons in the two energy bands.
\end{enumerate}
It follows from the above assumptions that the local equilibrium $g$ must be sought as the minimizer of the
Gibbs free-energy functional
\begin{equation}
\label{freenergy}
   \mathcal{E}(w) = \int_{\mR^6} \Tr\left\{ k_BT(  w \log w - w) + hw \right\} d\vp\,d\vx,
\end{equation}
among all positive-definite Wigner matrices $w$ sharing the macroscopic moments \eqref{momento_n} and
\eqref{momento_u}.
In \eqref{freenergy}, $k_B$ is the Boltzmann constant, $h$ is the matrix-valued symbol of the Hamiltonian (see \eqref{KpHamSymb}), and
$\log w$ is the matrix logarithm.
It can be shown \cite{BarlettiFrosali10} that the solution $g$ of such constrained minimization problem is
given by
\begin{equation}
\label{gform}
   g_\pm(\vx,\vp,t) = \e^{-\beta E_\pm(\vp) + \vB_\pm\cdot\vv_\pm(\vp) + A_\pm},
 \qquad
   \vctr{g}_\perp = 0,
\end{equation}
where $\beta = (k_BT)^{-1}$, and $A_\pm = A_\pm(\vx,t)$ and $\vB_\pm = \vB_\pm(\vx,t)$ are Lagrange
multipliers to be determined from the constraint equations
\begin{equation}
\begin{aligned}
&\int_{\mR^3}   g_\pm (\vx,\vp,t)\, d\vp =  n_\pm(\vx,t),
\\[6pt]
&\int_{\mR^3}  \vv_\pm(\vp)\, g_\pm (\vx,\vp,t) \,d\vp = n_\pm(\vx,t)  \vu_\pm(\vx,t).
\label{constraints}
\end{aligned}
\end{equation}
\par
Let us now assume that the time-scale over which the system is observed is much larger than the
relaxation time $\tau_c$ (the so-called hydrodynamic asymptotics).
In this limit, we have that $w \to g$ and we can rewrite Eq. \eqref{sum3} with $w_\pm = g_\pm$ and
${\vctr w}_\perp = {\vctr g}_\perp =   0$,
obtaining that the local equilibrium function satisfies
\begin{equation}
\label{sum4}
  \frac{\pt g_\pm}{\pt t }  + \vv_\pm  \cdot\nabla_\vx g_\pm  +   \vF \cdot \nabla_\vp  g_\pm    = 0.
\end{equation}
\begin{remark}
\rm
The quantum interference terms (i.e. the terms containing $\vctr{w}_\perp$ in Eq.\ \eqref{sum3}),
which are responsible for quantum coupling between the two bands \cite{Morandi09}, have disappeared in
our semiclassical hydrodynamic picture because $\vctr{g}_\perp = 0$.
When dealing with the semiclassical diffusive limit, however, we have to consider terms of order $\hbar$
in the semiclassical expansion of the quantum equilibrium (our $g$ is the leading order of such expansion)
and band-coupling interference terms appear \cite{BarlettiFrosali10,BarlettiMehats10}.
\hfill $\square$
\end{remark}
Integrating Eq.\ \eqref{sum4} over $\mR^3$, and using the constraints \eqref{constraints}, we have
\begin{equation}
\label{sum51}
\begin{aligned}
 &  \frac{\pt n_\pm}{\pt t }  +  \nabla_\vx \left(  n_\pm \momentou_\pm  \right)     = 0
\end{aligned}
\end{equation}
that is the continuity equation for $n_\pm$.
Multiplying  Eq.\ \eqref{sum4}  by $\vv_\pm$ and  integrating over $\vp$, we obtain the first-order
moment equation
\begin{equation}
\label{ueq}
\frac{\pt (  n_\pm \momentou_\pm)}{\pt t}
 + \nabla_\vx   \cdot \PP_\pm - \vF \cdot  \QQ_\pm =  0,
\end{equation}
that is the momentum balance equation,
where the tensors $\PP_\pm$ and  $\QQ_\pm$ are defined as follows:
\begin{equation}
\label{PQdef}
\PP_\pm = \int_{\mR^3} \vv_\pm  \otimes   \vv_\pm\, g_\pm\, d\vp,
\qquad
 \QQ_\pm = \int_{\mR^3}   (\nabla_\vp \otimes \vv_\pm) \,  g_\pm \,  d\vp.
\end{equation}
Recalling \eqref{velocity} and \eqref{effmass}, the tensor $\QQ_\pm$, which ``mediates'' the action of the force $\vF$,
can be written as
\begin{equation}
\label{Qalt}
 \QQ_\pm =  \int_{\mR^3}   (\nabla_\vp \otimes \nabla_\vp\,E_\pm) \,  g_\pm \,  d\vp
 = \int_{\mR^3}   \MM_\pm^{-1}(\vp) \,  g_\pm \,  d\vp,
\end{equation}
showing that $\QQ_\pm$ is the average inverse effective-mass.
For suitable values of $\valpha$ and $\gamma$, $\QQ_-$ can be negative: in this case
the lower-band electrons behave like positive-charged carriers (holes).
\par
We remark that the functions  $g_\pm$  have been determined by the maximum entropy principle and
depend implicitly on the moments $n_\pm$ and $\vu_\pm$ because the constraints \eqref{constraints}.
In this sense, the tensors  $\PP_\pm$ and  $\QQ_\pm$ can be regarded as functions of $n_\pm$ and $\vu_\pm$,
making the hydrodynamic system \eqref{sum51} + \eqref{ueq} formally closed.
\par
For future reference let us summarize here the hydrodynamic model that we have obtained: it consists of the
moment equations
\begin{equation}
\label{hydroeq}
\left\{
\begin{aligned}
&  \frac{\pt n_\pm}{\pt t }  +  \nabla_\vx \left(  n_\pm \vu_\pm  \right)     = 0,
\\[6pt]
&  \frac{\pt  (n_\pm \vu_\pm) }{\pt t}
 + \nabla_\vx   \cdot \PP_\pm  - \vF \cdot  \QQ_\pm = 0,
\end{aligned}
\right.
\end{equation}
and of the closure relations \eqref{PQdef} and \eqref{constraints}.
\section{The constraint equations}
\label{S5}
In this section we study the problem of how writing in a more explicit way the moment equations, that is expressing the
Lagrange multipliers $A$ and $\vB_\pm$, and consequently the tensors $\PP_\pm$ and  $\QQ_\pm$,
as functions of the moments $n_\pm$ and $\vu_\pm$.
\par
In order to simplify the notations we note that, both in the moment equations \eqref{hydroeq} and in the constraint
equations \eqref{constraints}, the $+$ and $-$ quantities are completely decoupled
(unless coupling mechanisms are introduced, as we will discuss in Section \ref{S6}).
Then, we can safely drop the $\pm$ labels everywhere, bearing in mind, however, that the $+$ and $-$ problems
are formally identical but physically different, because energies, velocities and effective-masses are different
in the two bands.
\par
In order to stress the dependence of the local-equilibrium on the Lagrange multipliers we put
\begin{equation}
  \phi(A,\vB,\vp) =  \e^{-\beta E(\vp) + \vB\cdot\vv(\vp) + A},
\end{equation}
and rewrite the constraint equations \eqref{constraints} as follows:
\begin{equation}
\label{newconstr}
\int_{\mR^3}  \phi(A,\vB,\vp)\, d\vp =  n,
\qquad
\int_{\mR^3}  \vv(\vp)\,  \phi(A,\vB,\vp)\, d\vp = n\vu,
\end{equation}
(recall that we are suppressing the labels $\pm$, and that $A$, $\vB$, $n$ and $\vu$ are functions of $(\vx,t)$).
Equations \eqref{newconstr} have to be regarded as a system of four scalar equations in the unknowns
$A$ and $\vB = (B_1,B_2,B_3)$, for given $n > 0$ and $\vu = (u_1,u_2,u_3) \in \mR^3$.
\par
Let us introduce the function $f(\vB)$ defined by
\begin{equation}
\label{fdef}
  \e^{f(\vB)} = \int_{\mR^3}   \e^{-\beta E(\vp) + \vB\cdot\vv(\vp)}  \, d\vp.
\end{equation}
By using
$$
   \vv(\vp)\, \phi(A,\vB,\vp) = \nabla_\vB \phi(A,\vB,\vp),
$$
we obtain that the constraint system \eqref{newconstr} is (formally) equivalent to
\begin{equation}
\label{syst}
\left\{
\begin{aligned}
  &\e^A \e^{f(\vB)} = n,
\\[4pt]
  &\nabla_\vB f(\vB) = \vu.
 \end{aligned}
 \right.
\end{equation}
From Eq.\ \eqref{syst} we see that $\vB$ only depends on $\vu$ and, once $\vB$ is solved from the second
equation as function of $\vu$, the remaining unknown $A$ is determined by $\e^A =  n\,\e^{-f(\vB)}$.
Moreover, using
$$
  \vv(\vp)\otimes\vv(\vp)\, \phi(A,\vB,\vp) = \nabla_\vB\otimes (\nabla_\vB \phi(A,\vB,\vp)),
$$
the tensor $\PP$ (see definition \eqref{PQdef}) can be written as
\[
\begin{aligned}
  \PP &= \e^A\,\int_{\mR^3} \nabla_\vB \otimes \left(  \nabla_\vB  \e^{-\beta E(\vp) + \vB\cdot\vv(\vp)}   \right) \, d\vp \\
  & =  \e^A\, \nabla_\vB \otimes \left(\nabla_\vB \e^{f(\vB)}\right)
  = \e^A \nabla_\vB  \otimes \left(  \e^{f(\vB)}  \left( \nabla_\vB  f(\vB)       \right)   \right) \\
  &= \e^A\e^{f(\vB)} \left[ \nabla_\vB  f(\vB) \otimes \nabla_\vB  f(\vB) +  \nabla_\vB \otimes\left( \nabla_\vB f(\vB) \right)\right]
\end{aligned}
\]
and therefore, using Eq.\ \eqref{syst},
\begin{equation}
\label{pressure}
  \PP = n\vu\otimes\vu + n \nabla_\vB \otimes (\nabla_\vB f(\vB)).
\end{equation}
This decomposition of $\PP$ shows that $\nabla_\vB \otimes (\nabla_\vB  f(\vB))$ plays the role of
pressure tensor in the Euler equations \eqref{hydroeq}.
Unfortunately, the ``mass'' tensor $\QQ$ has not a similarly simple expression in terms of $f(\vB)$.
\par
As already remarked, the form \eqref{syst} of the constraint equations allows to reduce the problem of the
solvability of $(A,\vB)$ as a function of $(n,\vu)$ to the solvability of $\vB$ as a function of $\vu$ from the
equation
$$
  \nabla_\vB f(\vB) = \vu,
$$
which is proven in the following theorem.
\begin{theorem}
\label{theo}
The mapping $\vB \in \mR^3 \mapsto \nabla_\vB f(\vB) \in  \mR^3$ is globally invertible.
\end{theorem}
{\it Proof.}
We first prove local invertibility.
Let $\vu(\vB) := \nabla_\vB f(\vB)$.
Using \eqref{pressure}, and recalling that $n>0$ is given, we have that
$$
   \frac{\pt u_i}{\pt B_j} = \frac{\pt^2 f}{\pt B_i\pt B_j} = \frac{\PP_{ij}}{n} - u_iu_j
$$ $$
   =  \frac{1}{n} \int_{\mR^3} (v_i(\vp) - u_i)(v_j(\vp) - u_j)\,\phi(A,\vB,\vp)\,d\vp,
$$
showing that the Jacobian matrix of the transformation is the covariance matrix
of $\vv(\vp)$, relative to the probability density $\phi(A,\vB,\vp)/n$, which is semi-definite positive.
The positive definiteness is readily proven by direct inspection, since
$$
  \sum_{i,j=1}^3  \frac{\pt u_i}{\pt B_j}\xi_i \xi_j =
   \frac{1}{n} \int_{\mR^3} \left[\vxi\cdot(\vv(\vp) - \vu)\right]^2 \phi(A,\vB,\vp)\,d\vp >0
$$
for every $\vxi \in \mR^3$ with $\vxi \not= 0$, which concludes the proof of local invertibility.
\par
In order to prove the global result, we resort to the classical result of Hadamard, that a local diffeomorphism
is global if an only if it is proper (the inverse image of a compact is compact).
In the present case this reduces to prove that, for every sequence $\vB_k \in \mR^3$ such that
$\abs{\vB_k} \to \infty$, also the image sequence $\vu_k = \vu(\vB_k) \in \mR^3$ is such that $\abs{\vu_k} \to \infty$.
Since $\abs{\vB_k} \to \infty$, we are interested in the asymptotic behavior of the distribution $\phi(A,\vB,\vp)$
for large $\abs{\vB}$.
Without loss of generality, we put here $m = 1$ and $\beta = 1$.
The critical points of  $\phi(A,\vB,\vp)$ (as a function of $\vp$) are determined by the condition
$$
  \nabla_\vp \left( E(\vp) - \vB\cdot\vv(\vp) \right) = 0.
$$
Recalling \eqref{Bands} and \eqref{velocity}, this leads to the condition
$$
  \vp \pm \nabla_\vp \abs{\vctr{h}(\vp)} - \vB \mp  \valpha\cdot\vB\, \nabla_\vp \nu_2(\vp) = 0,
$$
that is
$$
    \vp \pm \frac{(\valpha\cdot\vp)\,\valpha}{\left[ (\valpha\cdot\vp)^2 +\gamma^2 \right]^{1/2}}
     - \vB \mp \frac{(\valpha\cdot\vB)\,\valpha\,\gamma^2}{\left[ (\valpha\cdot\vp)^2 +\gamma^2 \right]^{3/2}} = 0.
 $$
Making the change of variable
$$
  \vq = \frac{\vp}{\abs{\vB}},
$$
we obtain the equation
$$
  \vq \pm \frac{(\valpha\cdot\vq)\,\valpha}{\abs{\vB} \left[ (\valpha\cdot\vq)^2 +\abs{\vB}^{-2}\gamma^2\right]^{1/2}}
  - \frac{\vB}{\abs{\vB}}
   \mp \frac{(\valpha\cdot\vB)\,\valpha\,\gamma^2}{\abs{\vB}^4 \left[ (\valpha\cdot\vq)^2 +\abs{\vB}^{-2}\gamma^2 \right]^{3/2}} = 0,
$$
which is asymptotically equivalent for $\abs{\vB} \to \infty$ to
$$
   \vq - \frac{\vB}{\abs{\vB}} = 0,
$$
i.e.\ to
$$
  \vp = \vB.
$$
Thus, we have shown that, for large $\abs{\vB}$, the distribution $\phi(A,\vB,\vp)$ has a single critical point
(which is clearly a maximum) at $\vp = \vB$.
Moreover, it decays like $\e^{-\abs{\vp}^2/2}$ away from the maximum.
This gaussian-like behavior ensures that
$$
   \frac{1}{n}\int_{\mR^3} \vp\, \phi(A,\vB,\vp)\,d\vp \sim \vB, \quad \text{as $\abs{\vB}\to \infty$.}
$$
Finally, since $\vv(\vp) = \vp \pm \nu_2(\vp)\valpha$, and $\nu_2(\vp)\valpha$ is a bounded quantity, we
also obtain
$$
 \vu =   \frac{1}{n} \int_{\mR^3} \vv(\vp)\, \phi(A,\vB,\vp)\,d\vp \sim
  \frac{1}{n}\int_{\mR^3} \vp\, \phi(A,\vB,\vp)\,d\vp \sim \vB,
$$
which shows that $\abs{\vu_k} \to \infty$ if $\abs{\vB_k} \to \infty$, concluding the proof.
\hfill $\square$
\section{Band coupling}
\label{S6}
As already remarked, the disappearance of the quantum interference terms in the semiclassical limit makes
our hydrodynamic model decoupled with respect to the two bands.
Coupling mechanisms can be introduced in two ways.
First of all, we may assume that the electric potential is composed of two parts:
$$
   V = V_\mathrm{ext} + V_\mathrm{int},
$$
where $V_\mathrm{ext}$ is the ``external'' part (taking account, e.g., of external bias, gate potentials,
and heterostructure potentials), while $V_\mathrm{int}$ is the ``internal'' (or self-consistent)
part, taking account of Coulomb repulsion between electrons.
In the simple mean-field model, this is given by the Poisson equation
\begin{equation}
\label{Poisson}
 \varepsilon_s \Delta V_\mathrm{int} = -q(n_+ + n_-),
\end{equation}
where $q$ is the elementary charge and $\varepsilon_s$ is the permittivity of the semiconductor.
The right-hand side depends on the total density $n_++n_-$, this coupling the upper-band and lower-band
populations.
\par
The other source of coupling derives from collisional mechanisms.
In order to introduce them, we have to go back to the kinetic level and add to the WBGK equation \eqref{WBGK1}
a suitable matrix-valued ``interband'' collisional operator $C(w)$ \cite{Rossani13}.
This is assumed to act on a much slower time scale with respect to $\tau_c$ (otherwise
it would affect the hydrodynamic  limit and destroy the structure of our MEP-based model).
Thus, we rewrite Eq.\ \eqref{WBGK1} with the (generic) additional terms:
\begin{equation}
\label{WBGK2}
\left\{
\begin{aligned}
 &\frac{\pt w_0}{\pt t}  + \frac{\vp}{m}\cdot\nabla_\vx w_0+ \vF \cdot \nabla_\vp w_0
 + \valpha\cdot\nabla_\vx w_2 =  \frac{g_0 - w_0}{\tau_c} + C_0(w) ,
\\[4pt]
 &\frac{\pt \vctr{w}}{ \pt t}+ \frac{\vp}{m}\!\cdot\!  \nabla_\vx  \vctr{w} + \vF \cdot \nabla_\vp \vctr{w}
  + \valpha\!\cdot\!\nabla_\vx w_0\,\vctr{e}_2 -  \frac{2}{\hbar}\, \vctr{h}(\vp) \!\times\! \vctr{w}
  =   \frac{\vctr{g} - \vctr{w}}{\tau_c} + \vctr{C}(w).
\end{aligned}
\right.
\end{equation}
Following the same arguments that led to Eq.\ \eqref{sum4}, we arrive at
\begin{equation}
\label{sum4bis}
  \frac{\pt g_\pm}{\pt t }  + v_\pm  \cdot\nabla_\vx g_\pm  +   \vF \cdot \nabla_\vp  g_\pm
 = C_\pm(g_+,g_-)
\end{equation}
(where we adopted a notation that stresses the fact that $g$ only depends on $g_+$ and $g_-$).
Taking the zeroth-order and first-order moments of this equation we get a modified version of
the hydrodynamic system \eqref{hydroeq}:
\begin{equation}
\label{hydroeq2}
\left\{
\begin{aligned}
&  \frac{\pt n_\pm}{\pt t }  +  \nabla_\vx \left(  n_\pm \vu_\pm  \right)  = N_\pm(n_+,n_-,\vu_+,\vu_-),
\\[6pt]
&  \frac{\pt  (n_\pm \vu_\pm) }{\pt t}
 + \nabla_\vx   \cdot \PP_\pm  - \vF \cdot  \QQ_\pm = \vU_\pm(n_+,n_-,\vu_+,\vu_-),
\end{aligned}
\right.
\end{equation}
where, of course,
\begin{equation}
\label{NUdef}
\begin{aligned}
 &N_\pm = \int_{\mR^3} C_\pm(g_+,g_-)\,d\vp
  \\[6pt]
  &\vU_\pm =  \int_{\mR^3} \vv_\pm(\vp)\,C_\pm(g_+,g_-)\,d\vp,
  \end{aligned}
\end{equation}
and  the dependence on $(n_+,n_-,\vu_+,\vu_-)$  follows from the MEP closure.
\par
Le us now list some possible choice of $C(w)$ in a simple BGK (relaxation time) form,
corresponding to different interband scattering mechanisms.
\paragraph{\it 1. Band-flip}
The electron undergoes a collision which exchange its band label from $+$ to $-$, or from $-$ to $+$.
Then we put
\begin{equation}
\label{sbf}
  C^\Bf(w) = - \frac{w - w_0\sigma_0}{\tau_\Bf} = - \frac{\vctr{w}\cdot\vctr{\sigma}}{\tau_\Bf}
\end{equation}
(where $\tau_\Bf$ denotes the characteristic time of band-flip scattering, which we assume constant for simplicity), so that
$$
  C^\Bf_\pm(w) = \mp \frac{w_+-w_-}{\tau_\Bf}
$$
(from which the band-flip is evident).
According to definition \eqref{NUdef}, therefore, we have
\begin{equation}
\label{NUbf}
  N^\Bf_\pm = \mp \frac{n_+-n_-}{\tau_\Bf},
  \qquad
  \vU^\Bf_\pm =  \mp \frac{n_+\vu_+-n_-\vu_-}{\tau_\Bf}.
\end{equation}
Note that the band flip mechanism conserves the total density and the momentum and relaxes the polarization
of density and momentum, (i.e.\ $n_+-n_-$ and $\vu_+-\vu_-$).
\paragraph{\it 2. Band relaxation}
An electron in the upper band undergoes a inelastic collision which scatters it to the lower band \cite{SIAP08}.
This mechanism is described by
\begin{equation}
\label{sbr}
  C^\Br(w) = - \frac{w_0\vctr{\nu} - \vctr{w}}{\tau_\Br}\cdot\vctr{\sigma},
\end{equation}
so that
$$
  C^\Bf_\pm(w) = \mp \frac{w_+}{\tau_\Br},
$$
(where $\tau_\Br$ denotes the characteristic time of band relaxation scattering, which we assume constant).
From definition \eqref{NUdef} we obtain
\begin{equation}
\label{NUbr}
  N^\Br_\pm = \mp \frac{n_+}{\tau_\Br},
  \qquad
  \vU^\Br_\pm =  \mp \frac{n_+\vu_+}{\tau_\Br}.
\end{equation}
Note that this mechanism conserves the total density an momentum and
depletes the upper band in favor of the lower.
\paragraph{\it 3. Isotropic interband scattering}
An electron undergoes a scattering event that changes its band label and re-distributes its momentum
according to a isotropic, thermal distribution.
This mechanism is described by
\begin{equation}
\label{sis}
  C^\Is(w) = - \frac{w - g^*}{\tau_\Is},
\end{equation}
where $\tau_\Is$ denotes the characteristic time of interband scattering, which we assume constant, and where $g^*$ is the isotropic version, with inverted densities,
of the MEP local equilibrium $g$, i.e.
\begin{equation}
\label{giso}
   g_\pm^*(\vx,\vp,t) = \frac{n_\mp}{z_\pm}\,\e^{-\beta E_\pm(\vp)},
 \qquad
   \vctr{g^*}_\perp = 0,
\end{equation}
where
\begin{equation}
 z_\pm = \int_{\mR^3} \e^{-\beta E_\pm(\vp)}\, d\vp,
\end{equation}
so that
\begin{equation*}
\begin{aligned}
\int_{\mR^3}   g_\pm^* (\vx,\vp,t)\, d\vp =  n_\mp(\vx,t),
\qquad
\int_{\mR^3}  \vv_\pm(\vp)\, g_\pm^* (\vx,\vp,t) \,d\vp = 0
\end{aligned}
\end{equation*}
(note the inverted band-labels of the density).
Then:
$$
  C^\Is_\pm(w) = - \frac{w_\pm-g^*_\pm}{\tau_\Is}
$$
and
\begin{equation}
\label{NUis}
  N^\Is_\pm = \mp \frac{n_+-n_-}{\tau_\Is},
  \qquad
  \vU^\Is_\pm =  - \frac{n_\pm\vu_\pm}{\tau_\Is},
\end{equation}

Note, therefore, that this scattering mechanism relaxes the current in both bands
and the density polarization .
%
%

%
%
\section{Conclusions}
We can finally summarize the hydrodynamic model emerged from our discussion.
It consists of the Euler-Poisson-like system
\begin{equation}
\label{hydroeq3}
\left\{
\begin{aligned}
&  \frac{\pt n_\pm}{\pt t }  +  \nabla_\vx \left(  n_\pm \vu_\pm  \right)  = N_\pm,
\\[6pt]
&  \frac{\pt  (n_\pm \vu_\pm) }{\pt t}
 + \nabla_\vx   \cdot \left(n\vu_\pm\otimes\vu_\pm +  n\TT_\pm \right)
 + \nabla_\vx \left(V_\mathrm{ext}+V_\mathrm{int}  \right) \cdot  \QQ_\pm = \vU_\pm,
\\[6pt]
 &\varepsilon_s \Delta V_\mathrm{int} = -q(n_+ + n_-),
\end{aligned}
\right.
\end{equation}
where:
$$
    N_\pm = N_\pm(n_+,n_-,\vu_+,\vu_-),
    \qquad
    \vU_\pm = \vU_\pm(n_+,n_-,\vu_+,\vu_-)
$$
are the coupling terms discussed above,
$$
  \TT_\pm = \nabla_{\vB_\pm}\otimes\nabla_{\vB_\pm} \log
  \int_{\mR^3}   \e^{-\beta E_\pm(\vp) + \vB_\pm\cdot\vv_\pm(\vp)}  \, d\vp
$$
is the pressure tensor, described in Sec.\ \ref{S5},
$$
 \QQ_\pm = \int_{\mR^3} \MM_\pm^{-1}(\vp) \,\e^{-\beta E_\pm(\vp) + \vB_\pm\cdot\vv_\pm(\vp) + A_\pm} d\vp,
$$
is the effective-mass tensor, also described in Sec.\ \ref{S5}, and the Lagrange multipliers $(A_\pm, \vB_\pm)$
can be uniquely solved as functions of the moments $(n_\pm,\vu_\pm)$ from the constraint equations
$$
\left\{
\begin{aligned}
   &\int_{\mR^3} \e^{-\beta E_\pm(\vp) + \vB_\pm\cdot\vv_\pm(\vp) + A_\pm} d\vp = n_\pm,
\\[6pt]
   &\int_{\mR^3} \vv_\pm(\vp) \,\e^{-\beta E_\pm(\vp) + \vB_\pm\cdot\vv_\pm(\vp) + A_\pm} d\vp= \vu_\pm,
\end{aligned}
\right.
$$
as proven in Theorem \ref{theo}.
%
%

\section*{Acknowledgements}
This work was dedicated to the memory of Prof. D. Ya. Petrina. The authors wish to recall the
many opportunities they had to meet him and appreciate his outstanding personality. Many meetings had place during his permanence in Italy and, mainly,
during his visits to our Department in Florence. Two of the authors had the chance to participate to the Conference ``Recent Trends in Kinetic Theory
and its Applications'' (Kyiv, Ukraine, 2004) and to experience his warm hospitality, of which they will treasure memory.

{\small
\renewcommand{\refname}{References}


\begin{thebibliography}{99}

\bibitem{Arnold96}
A.Arnold.  \emph{Self-consistent relaxation-time models in quantum mechanics}.
 Commun. Partial Differ. Equations \textbf{21(3-4)}, (1996) 473--506.

\bibitem{BarlettiBenAbdallah11}
L.Barletti, N. Ben Abdallah.
{\em Quantum transport in crystals: effective mass theorem and k.p Hamiltonians}.
Comm. Math. Phys. \textbf{307},  (2011) 567--607.

\bibitem{BarlettiFrosali10}
L.Barletti, G.Frosali. \emph{Diffusive limit of the two-band $\rm   \vk\cdot \vp$ model for semiconductors}.
J. Stat. Phys.  \textbf{139(2)},  (2010) 280--306.

\bibitem{BarlettiFrosali12}
L.Barletti, G.Frosali. \emph{Diffusive limits for a quantum transport model with a strong field}.
Transport Theory Statist. Phys.  \textbf{41(5-6)},  (2012) 473--493.

\bibitem{BarFroMor2014}
L. Barletti, G. Frosali, O. Morandi. \emph{Kinetic and Hydrodynamic Models for Multiband Quantum Transport in Crystals}.
In M. Ehrhardt and M. Koprucki (Eds.), ``Modern Mathematical Models and Numerical Techniques for Multiband Effective Mass Approximations'',  Lecture Notes in Computer Science, Engineering,
Springer, Berlin, 2014, pp. 1-49 (to appear)

\bibitem{BarlettiMehats10}
L. Barletti, F. M\'ehats, F, (2010).
\emph{Quantum drift-diffusion modeling of spin transport in nanostructures}.
J. Math. Phys.\ {\bf 51}, 053304 (2010).

\bibitem{SIAP08}
L.L. Bonilla, L. Barletti, M. Alvaro. \emph{ Nonlinear electron and spin transport
  in semiconductor superlattices}.
 SIAM J. Appl. Math. \textbf{69}(2), 494--513 (2008)

\bibitem{DegondRinghofer03}
P. Degond, C. Ringhofer. \emph{Quantum moment hydrodynamics and the entropy
 principle}.  J. Stat. Phys. \textbf{112}(3-4), 587--628 (2003)

\bibitem{Kane66}
E.O. Kane. \emph {The {k$\cdot$p} method}.
(In: Willardson, R.K., Beer, A.C. (eds.) Physics of {III-V} Compounds,
  Semiconductors and Semimetals), vol.~1, chap.~3. Academic Press, New York
  (1966)

 \bibitem{LaRosa09}
S. La Rosa, G. Mascali, V. Romano.
\emph{Exact maximum entropy closure of the hydrodynamical model for Si semiconductors:
the 8-moment case}.
 Siam J. Appl. Math. {\bf 70}(3), 710--734, 2009.


\bibitem{Morandi09}
O. Morandi.
\emph{Wigner-function formalism applied to the Zener band transition in a semiconductor}.
 Phys. Rev. B {\bf 80}, 024301(12) (2009).

\bibitem{Rossani13}
A. Rossani,
\emph{ Semiconductor spintronics in a participating phonon medium: Macroscopic equations}.
AIP Advances {\bf 3}, 092122 (2013). doi: 10.1063/1.4822161


\bibitem{Wenckebach99}
W.T. Wenckebach. \emph{ Essential of Semiconductor Physics}. J.Wiley \& Sons,
Chichester (1999)


\bibitem{Wigner32}
E. Wigner. \emph{ On the quantum correction for thermodynamic equilibrium}.
Phys. Rev. \textbf{40}, 749--759 (1932)

\bibitem{Wu97}
N. Wu. \emph{The Maximum Entropy Method}. Springer Verlag, Berlin (1997)

\bibitem{ZachosEtAl05}
C.K. Zachos, D.B. Fairlie, T.L. Curtright (eds.): \emph{Quantum mechanics in phase
  space}, \rm{World Scientific Series in 20th Century Physics}, vol.~34.
\newblock World Scientific Publishing Co. Pte. Ltd., Hackensack, NJ (2005).
\newblock An overview with selected papers


\end{thebibliography}
\end{document}